\RequirePackage{lineno}
\documentclass[jmp,superscriptaddress,showpacs,12pt,nofootinbib]{revtex4}

\usepackage{color,graphicx}
\usepackage[colorlinks=true]{hyperref}
\usepackage{amsfonts, amssymb, amsmath}

\def\be{\begin{eqnarray}}
\def\ee{\end{eqnarray}}
\def\bee{\begin{eqnarray*}}
\def\eee{\end{eqnarray*}}
 \def\pmx{\begin{pmatrix}}
 \def\emx{\end{pmatrix}}
 \def\bsq{\begin{subequations}}
\def\esq{\end{subequations}}

\newtheorem{thm}{Theorem}
\newtheorem{stmt}[thm]{Statement}

\newtheorem{cor}[thm]{Corollary}
\newtheorem{lemma}[thm]{Lemma}

\def\pf{\noindent {\bf Proof: }}
\newcommand{\norm}[1]{ \| #1  \|}

        \def\tr{\hbox{\rm Tr} \, }
         
          \def\ran{\hbox{\rm Range ~}}
          \def\trp{\hbox{\rm Tr} }
     
     \def\half{{\textstyle \frac{1}{2}}}
     \def\nn{\nonumber}

\def\cB{{\cal B}}

\def\cH{{\cal H}}
\def\cD{{\cal D}}

\def\cB{{\cal B}}
\def\cP{{\cal P}}
\def\cF{{\cal F}}
\def\cG{{\cal G}}
\def\cP{{\cal P}}

\def\bra{\langle}
\def\ket{\rangle}
\def\kb{ \ket \bra }

\def\rt2{ \frac{1}{\sqrt{2}} }

           \def\wtd{\widetilde}
\newcommand{\proj}[1]{ | #1 \kb  #1|}
\newcommand{\ovb}[1]{\overline{ #1 }}
   \def\qed{{\bf QED}}
\def\Raw{\Rightarrow}    

\def\ran{{\rm range}}

\def\eps{\epsilon}
\def\ot{\otimes}
\def\op{\oplus}

\begin{document}

%\ifdefined\lineno
%       \linenumbers
%\else
%\fi
%\linenumbers

\title{Comment on some results of Erdahl  and  the convex structure of reduced density matrices}

\author{Jianxin Chen}%
\affiliation{Department of Mathematics \& Statistics, University of
  Guelph, Guelph, Ontario, Canada}%
\affiliation{Institute for Quantum Computing, University of Waterloo,
  Waterloo, Ontario, Canada}%
\author{Zhengfeng Ji}%
\affiliation{Institute for Quantum Computing, University of Waterloo,
  Waterloo, Ontario, Canada}%
\affiliation{State Key Laboratory of Computer Science, Institute of Software, Chinese Academy of Sciences, Beijing, China}%
\author{Mary Beth Ruskai}%
\affiliation{Tufts University, Medford, MA
  02155, USA}%
  \affiliation{Institute for Quantum Computing, University of Waterloo,
  Waterloo, Ontario, Canada}%
\author{Bei Zeng}%
\affiliation{Department of Mathematics \& Statistics, University of
  Guelph, Guelph, Ontario, Canada}%
\affiliation{Institute for Quantum Computing, University of Waterloo,
  Waterloo, Ontario, Canada}%
\author{Duan-Lu Zhou}%
\affiliation{Beijing National Laboratory for Condensed Matter Physics,
  and Institute of Physics, Chinese Academy of Sciences, Beijing
  100190, China}%

%\today

\begin{abstract}
In {\em J. Math. Phys.}  {\bf 13},  1608--1621 (1972), Erdahl considered the convex structure of the set of $N$-representable 2-body reduced density matrices in the case of fermions. Some of these results have a straightforward extension to the $m$-body setting and to the more general quantum marginal problem.   We describe these extensions, but can not resolve a problem in the proof of Erdahl's claim that every extreme point is exposed in finite dimensions.   Nevertheless, we can show that when $2m \geq N$ every extreme point of  the set of $N$-representable $m$-body  reduced density matrices has a unique pre-image in both the symmetric and anti-symmetric setting. Moreover, this extends to the quantum marginal setting for a pair of complementary $m$-body and $(N-m)$-body  reduced density matrices.
 \end{abstract}
 
\maketitle
 
%  \linenumbers
 
\section{Introduction}  \label{sect:intro}
The development of quantum information theory has generated interest in what is known as the quantum marginal problem, i.e., the question of when a given set of reduced density matrices (RDMs) can be obtained from an $N$-body state.   A special case,  known as the $N$-representability problem asks when an $m$-body RDM for a fermionic system can be obtained from an $N$-body fermion state.    

This question was extensively studied in the 1960's and 1970's in the hope
of finding a way to use the $2$-body RDMs to simplify complex $N$-body
computations.    The $N$-representability of the 1-body RDM has a simple solution found independently by several groups, including \cite{Cole,Kuhn,GP,Yang:1962}. A 1-body RDM is $N$-representable if and only if its eigenvalues satisfy the condition $0 \leq  \lambda_k \leq \tfrac{1}{N} $ which expresses the Pauli exclusion principle. 

However, $N$-representability for the 2-body RDM is a far more challenging problem on which little progress was made for over 30 years.   In 2007, the hardness of this problem was formally recognized by proving that it belongs to the complexity class known as QMA-complete \cite{QMA}, which is the quantum analogue of NP-complete, i.e. testing for $N$-representability would require exponential time  even on a quantum computer in the worst cases.

Slightly earlier, Klyachko \cite{Kly,KlyCMP}  found a complete solution of the pure state $N$-representability problem for the 1-body RDM.   The earlier solution of the mixed state problem can be restated as the fact that the extreme points of the convex set of 1-body RDMs arise from $N$-body Slater determinants. However, little was known about the images of more complex states beyond an abstract induction result of Coleman \cite{Cole} which can be viewed as a constrained version of Weyl's problem \cite{MBR}.
  
 These advances came as a result of recent work in quantum information theory in which a RDM is often called a ``quantum marginal''.   There are several variants \cite{CM,Kly0} of the question of whether of  a given set of quantum marginals is consistent; the $N$-representability problem can be regarded as a special case of one of these. Although consideration of symmetry constraints  is at the heart of $N$-representability, some fundamental aspects of the convex structure carry over from the anti-symmetric $N$-representability situation to the more general quantum marginal setting for $m$-body RDMs.  

 In \cite[Section 3]{E},  Erdahl claims that in finite dimensions, every extreme point of the convex set of $N$-representable reduced density matrices (RDMs) is exposed.  In trying to extend  this argument to quantum marginals, we found a subtle problem in the proof, which we describe in Section~\ref{sect:glitch}. Because some of the results in \cite{E} easily extend to quantum marginals, we introduce notation in Section~\ref{sect:not}  which makes such generalizations transparent. 
         
However, Erdahl's primary application of this result, i.e., his proof \cite[Section 6]{E}  that when $2m \geq N$, the pre-image of an exposed point of the set of $N$-representable fermionic $m$-body RDMs  is unique holds for arbitrary extreme  points. One needs only the observation that the pre-image of a face is a face.
         
In addition we  use one of his key observations \cite[lemma~6.1]{E} to show that a pair of complementary quantum marginals has a unique pre-image when both are extreme points of their respective sets of RDMs.  This is a rather remarkable property of extreme points because, in general, a pair of density matrices whose Hilbert spaces have a non-trivial overlap seems to be essential to the unique determination of a pure state pre-image.     Moreover, this contrasts sharply with the situation when $2m < N$ for which a single extreme $m$-body RDM can have multiple pre-images as shown recently in \cite{Ocketal}.
          
The rest of this paper is organized as follows. Section~\ref{sect:not} introduces necessary concepts and notation for both anti-symmetric case and quantum marginals.  We then describe the problem in Erdahl's proof of the claim that every extreme point is exposed in finite dimensions in Section~\ref{sect:glitch}. Inspired by Erdahl's primary application in  \cite[Section 6]{E} that when $2m \geq N$ every exposed point of the set of $N$-representable RDM has a unique pre-image, we study the pre-images of extreme points and interior points in Section~\ref{sect:uniq}. Specifically, before describing our main results of Section~\ref{sect:uniq}, we first give some intuition and background in Sub-Section~\ref{sub:intro}. Then, by applying a technical lemma provided in Sub-Section~\ref{sub:key}, we show that Erdahl's result holds for extreme points even if they are not exposed in Sub-Section~\ref{sub:ext}. In Sub-Section~\ref{sub:mix}, we show that if the contraction map has a non-trivial kernel, then every interior point of the set of $N$-representable RDM has non-unique pre-images.  Conclusion and future work can be found in Section~\ref{sect:sum}.

% \pagebreak
                
\section{Notation}   \label{sect:not}

\subsection{The antisymmetric case}   \label{sect:anti}

Let $\cH$ be a $d$-dimensional Hilbert space and let  $\cH_N \equiv
\cH^{\ot N}_{-} = \cH^{\wedge N} $ be the anti-symmetric subspace of
its $N$-fold tensor product.  Let $\cD(X)$ denote the set of density matrices for an antisymmetric subspace $X \subseteq  \cH^{\ot N}$, i.e., 
\be
  \cD(X) = \{ \rho \in \cB(X) : \rho \geq 0, ~~ \tr \rho = 1 \}
\ee     
where $\cB(X)$ is the set of all bounded linear operators on $X$.

Let $\cD^m(X)$ denote the corresponding set of reduced density matrix (RDM), i.e.,  
\be
   \cD^m(X) &  \equiv & \{ \rho_m \in \cD(\cH^{\otimes m})       : 
     \exists  ~ \rho \in \cD(X) ~ \hbox{with}  ~ \tr_{m+1, \ldots, N} \rho \,  = \rho_m \}  \nn  \\
     & = & \hbox{convex hull} ~  \{ \tr_{m+1, \ldots, N} \proj{\psi} : |\psi\ket \in X \}.
\ee
Then $ \cD^m(\cH^{\wedge N} )$ is the set of $N$-representable RDM.   The polar cone is
\be
  \cP \big[  \cD^m(X) \big] & \equiv &
     \{ V  \in \cB(\cH^{\ot m}) : \tr (V \rho_m) \geq 0 ~~ \forall ~ \rho_m \in  \cD^m(X) \}.
\ee
Every $V \in  \cP \big[  \cD^m(X) \big]  $ can be associated with a positive
semi-definite Hamiltonian in $\cB(X) $ given by
\be
   H_N(V) = \sum_{j_1, j_2, \ldots , j_m }  V(j_1, j_2 \ldots , j_m ) 
\ee
 where  $V(j_1, j_2 \ldots , j_m ) $ acts on $\cH_{j_1} \ot \cH_{j_2} \ot \ldots \ot \cH_{j_N} $.
 An extreme point is exposed if there is a $V \in   \cP \big[  \cD^m(\cH^{\wedge N}  )\big] $
 such that $\tr  (V \rho_m) = 0$ and $\tr (V \gamma_m) > 0$ for all $\gamma_m \neq \rho_m$.
 It then follows immediately from the variational principle that the pre-image of $\rho_m$ is a density matrix for the ground state eigenspace of $ H_N(V) $.  (Although the generic situation is a one-dimensional eigenspace, it was shown in \cite{Ocketal} that the pre-image of an exposed point can  be the set of density matrices associated with a degenerate ground state eigenspace of an $m$-local Hamiltonian.)
 
%A face of a convex set $C$ ($C=\cD^m(\cH^{\wedge N} )$ here) is defined as a convex subset $F$ of $C$ such that, for any line segment  $L\subseteq C$, if $L$ intersects $F$ at some point other than the two end points of $L$, then $L\subseteq F$. An exposed face of a convex set $C$ is defined as the intersection of $C$ and a certain non-trivial supporting hyperplane to $C$~\cite{Roc96}.
%
%An open subset of a closed set $C$ is the intersection of $C$ and some open set of the full metric space in which $C$ is embedded.

Let  $\{ \rho_{\lambda} = \lambda \rho + (1-\lambda)  \rho^\prime : \lambda \in [0,1] \} \subset {\cal C} $ denote a line segment of a convex set  ${\cal C}$.  A face $\cF$ of ${\cal C}$ is a subset with the  following property.   Whenever some element  $\rho_{\lambda} $ of a line segment  is in $\cF$ for some $\lambda \in (0,1) $, then the entire line segment is in $\cF$. An extreme point of ${\cal C}$ is a face of dimension $0$.  A face $ \cF $ of  $ \cD^m(X) $ is exposed if there is an element  $V$ of the polar cone such that   $\tr  (V \rho_m) = 0 ~~ \forall \rho_m \in \cF$ and $\tr  (V \gamma_m) > 0$ for all $\gamma_m \notin \cF$.

Note that if a density matrix $\rho \in \cD(X) $ does not have full rank, then it is in the face $(\ker \rho)^\perp  \cap   \cD(X) $, i.e., the set of all density matrices $\gamma$ for which range $\gamma \subseteq$ range $ \rho$. Therefore, any state in the interior of $ \cD(X) $ must have full rank.

%\bigskip
 %  % \pagebreak
 %\bigskip
 
 \subsection{Quantum marginals}   \label{sect:qmarg}

Some results about the convex structure of the set of $N$-representable RDMs
have straightforward extensions to convex sets associated with quantum marginal problem. In this situation  we want to know if there is an $N$-body state consistent with a set of $m$-body RDMs.

In the general case, we can replace $\cH^{\wedge N}  $ by an arbitrary subspace
$\cH_N \subseteq \cH^{\ot N}$.    For 
a given set of indices  $J = \{ j_1, j_2, \ldots, j_m \} $, let $J^C$ denote the 
complement in $\{ 1,2, \ldots , N \} $ and for any $N$-body state   
$\rho_{1,2 ,\ldots ,N}   \in  \cD(\cH_N ) $ define
\be
    \rho_{j_1 j_2 \ldots j_m} \equiv \rho_J = \trp_{J^C} \rho_{1,2 ,\ldots ,N} .
\ee
What we called  $\rho_m$ in the previous section is  more properly called
 $\rho_{1,2, \ldots ,m} $. When  $\cH_N$ is the symmetric or anti-symmetric subspace of $\cH^{\ot N}$ this slight abuse of notation is justified by the fact that  $\rho_m$ determines $\rho_{j_1 j_2 \ldots j_m}$ for any set of indices.   If we let  $\cH_J  \subseteq \cH^{\ot m} $ 
denote the subspace induced  by $H_N$, then, as above for any subspace $X \subseteq \cH_N$ we define
\be     
      \cD_J(X)  & \equiv & \{  \rho_J \in \cD(\cH_J) : \exists ~~ \rho_{1, 2,        \ldots ,N} \in \cD(X)
       \hbox{ such that } \rho_J =  \trp_{J^C} \rho_{1, 2, \ldots ,N}  \} .
\ee 

 We can then replace $\rho_m$  by a vector, e.g.,
\bee 
      \vec{R}_2    = (\rho_{12}, \rho_{13}, \ldots  ,\rho_{1n}, \rho_{23}, \ldots  ,\rho_{N-1,N} ) 
\eee
    or, more generally
\bee
     \vec{R}_m   = (\rho_{1,2, \ldots ,m} ,  ~ \rho_{1,2, \ldots ,m-1, m+1} , ~  \ldots , ~ \rho_{N-m+1, \ldots, N-1, N} ) \nn 
\eee
consisting of all possible $m$-body RDMs in some prescribed order.  
Then we define
 \be 
 \cD^m(X) &  \equiv  & \{      \vec{R}_m     \in \cD(X)        : 
     \exists  ~ \rho \in \cD(X) ~ \hbox{with}  ~ \rho_J  \in   \cD_J(X)  ~~ \forall  J  \hbox{ with } |J| = m \} .
 \ee
Given a vector $\vec{R}_m$ with elements in $\cD(\cH^{\ot m} ) $ the consistency problem for quantum marginals asks if there is a state $ \rho_{1,2, \ldots ,N}   \in  \cD(\cH_N ) $  
whose RDMs are given by the elements of $\vec{R}_m$.   
When  $\cH_N = \cH^{\wedge N} $  this is the $N$-representability problem. Both $ \cD_J(X) $ and $ \cD^m(X) $ are closed convex sets.

The polar cone $\cP\big[\cD^m(\cH_N)\big] $ consists of vectors
 \be
     \vec{V}_m =  (V_{1,2, \ldots ,m} ,  ~ V_{1,2, \ldots ,m-1, m+1} , ~  \ldots , V_{N-m+1, \ldots, N-1, N} )
 \ee
with elements  $V_{j_1 j_2 \ldots j_m}  \in \cB( \cH_{j_1} \ot \cH_{j_2} \ot \ldots \ot \cH_{j_m}) $
and  $\tr (\vec{V}_m \cdot \vec{R}_m )\equiv \sum_{|J| = m }   (\tr V_{J} \rho_{J}) $.  Elements of the polar cone of  $\cD(\cH^{\ot m} ) $ are associated with $m$-local $N$-body Hamiltonians $H_N(V) = \sum_{|J| = m }  V_{J} $ which are positive semi-definite; the extreme rays have a ground state with eigenvalue zero.  

We may and do carry over the descriptions of face and exposed face using the concept of polar cone to the quantum marginal case.

%
% We will have occasion to consider complementary pairs of quantum marginals
% $( \rho_{J}, \rho_{J^C} )$.   For fixed $J$ and $N$ the  
%set 
%\be  \label{pairdef}
%   \cD_{J,J^C}(X) = \{ ( \rho_{J}, \rho_{J^C} ) :  \rho_{J} \in  \cD_J(X)  \text{and}  \rho_{J^C} \in   \cD_{J^C}(X)
%\ee
%is also a closed convex set.    
% 
 
% \pagebreak
 
\section{A subtle problem in Erdahl's lemma} \label{sect:glitch}

% \subsection{The problem}

We now describe the problem in Erdahl's proof of the claim that every
extreme point is exposed in finite dimensions.   For simplicity, we consider
the original anti-symmetric situation with $\cH_N =  \cH^{\wedge N} $ and use the notation introduced in Section~\ref{sect:anti}.
However, it should be clear that everything goes through in the more
general quantum marginal setting with $\rho_m$ replaced by $ \vec{R}_m$.
 
Let $A,B$ be subspaces with $\cH^{\wedge N} \supset A \supset B$, and 
suppose that  $\cF =  \cD^m(A)$ is an exposed face of $\cD^m(\cH^{\wedge N})$ the set of $N$-representable RDM  and $\cG =  \cD^m(B) $ is another face which lies in $\cF$.  In \cite[section 3]{E}\footnote{It is ironic that there is a typesetting error in the title of this section so that what was intended as ``In Finite Dimensions \ldots '' appears as ``INFINITE DIMENSIONS ... ''  rather than ``IN ~FINITE DIMENSIONS''. } Erdahl states the following as   Lemma~3.1.     
%  \bigskip   %\bigskip    % \pagebreak 
 \begin{stmt}
  If   $\cG =  \cD^m(B) $  is an exposed face of the convex set $ \cD^m(A)$,
then it is also an exposed face of  $ \cD^m(\cH^{\wedge N}) $.
 \end{stmt}
By assumption, 
\begin{itemize}
\item[a)]  there is a $V  \in   \cP \big[ \cD^m(\cH^{\wedge N}  )\big] $ which exposes $\cF$, i.e.,
$\tr (V \rho_m) = 0 ~~ \forall ~ \rho_m \in \cF$ and 
$\tr (V \rho_m) > 0  ~~ \forall ~ \rho_m \in  \cD^m(A) \backslash \cF$.
\item[b)]    there is a $W  \in \cP \big[ \cD^m(A) \big] $ such that 
$\tr (W \rho_m) = 0 ~~ \forall ~ \rho_m \in \cG$ and 
$\tr (W \rho_m) > 0  ~~ \forall ~ \rho_m \in  \cD^m(A) \backslash \cG$. 
\end{itemize}

Although $H_N(V) $ is positive semi-definite on $\cH^{\wedge N}$, the operator  $H_N(W) $ need not be positive semi-definite. 
However, Erdahl claims that one can find a $t > 0 $ such that
$t H_N(V) + H_N(W) $ is positive semi-definite. But we  claim that  this is true if and only if $B \subseteq \ker  H_N(W)  $.
(Erdahl states that  ``clearly'' $B \subseteq \ker  H_N(W)  $, but gives no proof.)

We can decompose  
\bee 
    \cH    =\cH_1 \op \cH_2 \op \cH_3  \equiv 
    B \op \left(A \cap B^\perp\right) \op A^\perp,  
\eee
and write 
$H_N(V) $ and $ H_N(W) $ as block matrices accordingly
\be
       H_N(V)  = \pmx 0 & 0 & 0 \\   0 & 0 & 0 \\   0 & 0 & X_{33} \emx, \qquad
         H_N(W)  = \pmx 0 & 0 & Y_{13} \\   0 & Y_{22} & Y_{23} \\   Y_{31} & Y_{32}  & Y_{33} \emx .
\ee
The matrices  $X_{33}$ and $Y_{22}$ are strictly positive  definite by construction. However,  the $2 \times 2$   block submatrix of $ t H_N(V) + H_N(W) $ obtained by omitting the second row and column is
\be
   \pmx 0 &   Y_{13} \\  Y_{31} &     t X_{33} +  Y_{33}  \emx  
\ee
which can {\em never be positive semi-definite for any $t > 0$ unless} $Y_{13} = 0 $, which is equivalent to  $B \subseteq \ker H_N(W) $.   Although
\bee
 | \psi \ket \in B ~~ \Raw  ~~ \trp_{1 ,\ldots, m}  \,  W   \big( \trp_{m+1, \ldots ,N} \proj{\psi} \big) = 0 
     ~~ \Raw  ~~  \bra \psi, H_N(W) \, \psi \ket = 0,
 \eee 
this does not imply that $| \psi \ket \in \ker H_N(W) $ because 
$H_N(W) $  need not be positive semi-definite~\footnote{Indeed, if $H_N(W) $ is not positive semi-definite, then there will always 
be vectors in $\cH^{\wedge N} $ such that $\bra \psi,  H_N(W)  \psi \ket = 0$ but $\psi \notin \ker  H_N(W) $.  This is because on  $H_N(W) $ will have both positive and negative eigenvalues on $[\ker H_N(W)]^\perp$ which implies that $0$ in the numerical range of  $H_N(W) \big|_{[\ker H_N(W)]^\perp}$.}.  We can only conclude that  $| \psi \ket \in \ker  P_A H_N(W) P_A $ where $P_A $ is the projection onto $A = \ker H_N(V) $.    In general, we do not expect that $P_A H_N(W) P_A $ is an $m$-body Hamiltonian.

If it were true that $Y_{13} = 0 $, we could simplify Erdahl's argument slightly because it suffices to consider the non-zero $2 \times 2$ submatrix of   
\be   \label{red22}
 t H_N(V) +  H_N(W) =  
    \pmx 0 & 0 & 0 \\   0 &  Y_{22} & Y_{23} \\   0 & Y_{32}  &  t X_{33} + Y_{33} \emx  .
      % =    \pmx 0 & 0 & 0 \\   0 &   \eps Y_{22} &  \eps Y_{23} \\   0 & \eps Y_{32}  &   X_{33} + \eps Y_{33} \emx  
\ee  

As noted above, the assumptions on the regions for which $\tr (V \rho_m) > 0 $ and $\tr (W \rho_m) > 0$ imply that both $X_{33} $ and $Y_{22} $ are positive definite.  Therefore, using a standard result for $ 2 \times  2$  block matrices, \eqref{red22} is positive semi-definite if and only if 
\be      \label{poscond}
            Y_{23}^{\dagger} Y_{22}^{-1} Y_{23} \leq   t X_{33} +  Y_{33}  
\ee
which always holds for $t$ sufficiently large.   To be precise, we could find $\mu, \lambda > 0 $ such that  $Y_{22} \geq \mu I $ and $X_{33} \geq \lambda I$ so that it suffices to choose  
\be
    t > \tfrac{1}{\mu \lambda } \big( \norm{ Y_{23} }^2 - \mu \norm{ Y_{33}  } \big).
\ee

\section{Pure state pre-images and the uniqueness question}  \label{sect:uniq}

\subsection{Introduction and Intuition}\label{sub:intro}

In \cite[Section 6]{E} Erdahl proves that when $2m \geq N$ every exposed point of the set of $N$- representable RDM has a unique pre-image.  We show that this holds for extreme points even if they are not exposed by showing that \cite[Lemma~6.1]{E} still holds in the form of Theorem~\ref{thm:NmON} below. Remarkably, we use this result to extend Erdahl's  result to complementary pairs of quantum marginals, even without permutational symmetry or overlap.

\begin{thm}   \label{thm:unique}
If $2m \geq  N$  any extreme point of the set of $N$-representable RDM
has a unique pre-image.  Moreover, for fixed $J$  whenever both $  \rho_J$ and $\rho_{J^C}$ are extreme points of  $\cD_J(\cH_N) $ and $\cD_{J^C}(\cH_N) $ respectively, then the pre-image of the pair  $(  \rho_J, \rho_{J^C}) $ is unique.
\end{thm}

The proof of Theorem~\ref{thm:unique} is an immediate consequence of
Theorem~\ref{thm:NmON}  which is a straightforward generalization of Lemma~6.1 of Erdahl \cite{E}.    Before describing these results, we give some intuition and background.

%The intuition underlying this result comes from the so-called Schmidt
%decomposition in the special case that  $\rho_m$ has non-degenerate eigenvalues with eigenvectors $\{|\chi_k \ket \}$. When the pre-image is anti-symmetric and $m \geq N-m$, we can find $\rho_{N-m}$ and its eigenvectors $|\phi_k \ket$.   Then any pure state pre-image
%must have the form
%\be  \label{Schmidt}
%     | \Psi \ket   =  \sum_k \,  \mu_k e^{i\theta_k} \,  | \chi_k  \ot  \phi_k \ket.
%\ee
% If $|\Psi\rangle$ is not the unique pure state pre-image, it must have the form
% \be
%       | \Psi \ket   = x  |\psi_1 \ket +  y |\psi_2 \ket  \equiv 
%            \sum_{k \in K_1} \,  \mu_k e^{i\theta_k}\,   | \chi_k  \ot  \phi_k \ket
%             +   \sum_{k \in K_2} \,  \mu_k e^{i\theta_k}\,   | \chi_k  \ot  \phi_k \ket
% \ee
%with $K_1, K_2 $ disjoint and non-empty, $x, y$ non-zero  and $|x|^2 + |y|^2 = 1$.  Since the  $\{\phi_k\}$
%are orthogonal
%\be
%    \rho_m = |x|^2 \trp_{m+1, \ldots, N } \, \proj{\psi_1} + |y|^2  \trp_{m+1, \ldots, N }  \proj{\psi_2}.
%\ee
%This is a mixture and hence not extreme.  
The intuition underlying this result comes from the so-called Schmidt
decomposition in the special case that  $\rho_m$
has non-degenerate eigenvalues with eigenvectors $\{|\chi_k \ket\} $.
When the pre-image is anti-symmetric and $m \geq N-m$, we can find
$\rho_{N-m}$ and its eigenvectors $\{|\phi_k \ket\}$.   Then any pure state pre-image
must have the form
\be  \label{Schmidt}
    | \Psi \ket   =  \sum_k \,  \mu_k \, e^{i \theta_k} | \chi_k  \ot  \phi_k \ket.
\ee
If $|\Psi\ket$ is not unique (up to an overall phase factor), then, after suitable absorption of
some phase factors into  the  $| \chi_k \ket$,  one can find a pair of pre-images
which can be written as non-trivial  superpositions  of the form
\be
      | \Psi_j \ket   = x  |\psi_1 \ket + e^{i \omega_j} y |\psi_2 \ket  \equiv 
           \sum_{k \in K_1} \,  \mu_k \,   | \chi_k  \ot  \phi_k \ket
            +  e^{i \omega_j} \sum_{k \in K_2} \,  \mu_k \,   | \chi_k  \ot  \phi_k \ket \textrm{\ \ \ \ } j=1,2
\ee
with $K_1, K_2 $ non-empty and disjoint,  $|x|^2 + |y|^2 = 1$ and  
$0\leq \omega_1 \neq \omega_2 < 2 \pi $. 
Since $\cH_N$ is a vector space, both  $ | \Psi_1 \ket   -  | \Psi_2 \ket $
and $e^{- i \omega_1} | \Psi_1 \ket   - e^{- i \omega_2}  | \Psi_2 \ket  $
are in $\cH_N$ which implies that both $ |\psi_1 \ket $ and $ |\psi_2 \ket $
are in $\cH_N$.   Then
using the orthogonality of the  $| \phi_k \ket $,
we find
\be
   \rho_m = |x|^2 \trp_{m+1, \ldots ,N } \, \proj{\psi_1} + |y|^2  \trp_{m+1, \ldots ,N }  \proj{\psi_2}.
\ee
This is a mixture and hence not extreme.  This argument is essentially due to D. Smith \cite{smith} who considered the case $m = N-1 $ and for $m = 2, N = 3$ even analyzed the general case with  degenerate eigenvalues.  This argument is also the basis for a result of Diosi \cite{diosi} which we state   next.

\begin{thm}  \label{thm:diosi} 
{\rm (Diosi-Smith)}   Almost every pure state $|\psi \ket  \in \cH^{\ot N} $ is uniquely determined by a pair of RDM  $(\rho_J, \rho_{J^\prime} )$ with 
$ J \cap J^\prime \neq \O$ and $J \cup J^\prime = \{ 1, 2, \ldots N \}$.
\end{thm} 
\noindent{\bf Sketch of proof :}   For $N \geq 3$,  the set of $|\psi\ket$ for which $\rho_J$ has degenerate eigenvalues has measure zero.   Moreover, the hypotheses of the theorem imply that $J^C   \subset J^\prime$ so that we can determined $\rho_{J^C}$  by taking a suitable partial trace of $\rho_{J'}$.  Therefore, we can assume that we have an expansion as in \eqref{Schmidt}.  Since we also have $ (J^\prime)^C  \subseteq J$
the eigenvectors of $ \rho_{J^\prime} $ and $ \rho_{(J^\prime)^C}$ give  a second expansion.  Equating these two expansions pointwise gives a set of linear equations \footnote{In general, an arbitrary pair of RDM $\rho_J$ and $\rho_{J^c}$, will not even have a consistent set of linear equations  
with a solution for  $x_j$ and $x_j^\prime$, much less a set with unit norm.} for $x_j \equiv e^{ i \theta_j }$ and   $x_j^\prime \equiv e^{ i \theta_j^\prime }$.  We have at most $\dim \cH_J + \dim \cH_{J^C} $ unknowns and $\dim \cH_N$ equations.   In typical situations  $\dim \cH_N \approx  \dim \cH_J \dim \cH_{J^C} $. When this is not true, as in the antisymmetric case, the symmetry restrictions give additional equations.  Thus, in general, we expect a  unique  solution for $x_j $ and $x_k^\prime$. With the additional requirement that acceptable solutions must satisfy $| x_j | = |x_k^\prime | = 1$, the situations in which multiple solutions exist for the phases $\omega_j $ will be expected to be very rare.     \qed

Theorem~\ref{thm:unique} says that when both  $\rho_J$ and $\rho_{J^C}  $  are extreme points of their respective set of RDMs, then
the pair $(\rho_J, \rho_{J^\prime})$  determines a unique pure pre-image $| \psi \ket $ although  for  $ J^\prime = J^C $,  $  J \cap J^\prime  = \O$ precluding a second expansion. In view of the argument above, the claimed result  might seem too good to be true. However, when $\cH_N = \cH^{\ot N} $ with no constraints, 
$\cD_J(\cH^{\ot N} ) =  \cD(\cH_{j_1} \ot \cH_{j_1}  \ot \ldots \ot\cH_{j_m} )$
which is simply the convex hull of projections onto pure states $|\chi\rangle \in \cH^{\ot m}$.  Therefore, in the complete absence of constraints,  Theorem~\ref{thm:unique} simply reflects the fact that the extreme pairs
have the form  $ \proj{\chi}, \proj{\phi} $ and the pre-image $|\psi \ket = | \chi \ot \phi \ket $ is a pure product state.  On the other hand,  when $\cH_N$ is the symmetric or anti-symmetric subspace of $\cH^{\ot N} $ and 
 $2m = N$, we also have $J = \{ 1,2 \ldots m \} $ and $J^\prime = J^C = \{ m+1, m+2 , \ldots ,N \}$. However, permutations of the single expansion \eqref{Schmidt} which exchange $j \leftrightarrow k$ with $j \leq m$ and $k > m$ give linear equations which determine the phases, as above.
 
More generally, one might have other symmetry constraints on $\cH_N$, e.g., rotational symmetry or translational symmetry for spin lattices.   In such situations, the equations for determining the phases might not be as transparent as for permutational symmetry.  Theorem~\ref{thm:unique} says that, nevertheless, the condition of being an extreme point of 
$\cD_{J,J^C} (\cH_N) $ is sufficiently strong to uniquely determine the pre-image.   To prove this, we need some additional lemmas.

\subsection{Key Lemmas} \label{sub:key}

\begin{lemma}    \label{lemma:key}
Let  $\cG$ be the span of the extreme points of a face of  $\cD(\cH_N)$  for which every  state maps to a unique $m$-body RDM, i.e.
\be
 \rho_{J} = \trp_{J^C}  \proj{\psi}  \qquad \forall ~~  |\psi \ket   \in \cG .
\ee
%, i.e.   for symmetric 
% or anti-symmetric states
%  $\trp_{m+1, \ldots N} \proj{\psi}  = \gamma_m$ for all $\psi \in \cG$ 
Then for any pair of orthogonal vectors $| \psi_j \ket, |\psi_k \ket $ in $\cG$,

{\em (i)}   The  $(N \! - \! m)$-th order transition density matrix  $\trp_{J^C}  |\psi_j \kb \psi_k | = 0$; 
\newline   (in particular, for  $| \psi \ket $ symmetric  or anti-symmetric  
    $ \trp_{m+1, \ldots N} |\psi_j \kb \psi_k | = 0$).  
      
{\em  (ii)}  For any $m$-body operator $B_{j_1, \ldots ,j_m } = B_J $ acting on
       $\cH_{j_1} \ot  \cH_{j_2} \ot \ldots  \ot \cH_{j_m}$,  % we have 
       $\bra \psi_j,  B_{J} \psi_k \ket = 0$.
\end{lemma}
 
%     
%     \begin{cor}
%   The subspace $\cG$  defines a quantum code which
%   can correct $\lfloor  m/2 \rfloor $ errors, i.e. $\nu$ errors when $m = 2 \nu$ is even.
%   When $m = 2 \nu+1$ is odd, the code can correct $\nu$ errors
%   and detect $\nu+1$ errors.
%       \end{cor}
%      \noindent{\bf Proof:}  Since all $\psi \in \cG$ have the same
%      $m$-body RDM, for any O.N. basis of $\cG$ and $m$-body operator,  $\bra \psi_k ,B_m \psi_k \ket$
%      is independent of $k$.    Orthogonality for $j \neq k$ is precisely (ii).

 \noindent{\bf Proof :} When $\dim \cG = 1$, the result holds trivially, since there are no orthogonal pairs of vectors.  Therefore, we assume that $\dim \cG \geq 2$. Let $\{ \rho_{J} \} $ be the unique set  of $m$-body RDM onto which all normalized vectors in $\cG$ are mapped.   Then for any $a, b > 0$ with $a^2 = b^2 = 1$ and any $\theta$, consider the RDM 
 $\rho_{J}$ of $ a |\psi_1 \ket + e^{i \theta} b  |\psi_2 \ket$:
\be
      \rho_{J} & = &  \trp_{J ^C} \, \Bigl[ (a |\psi_1\ket + e^{i \theta} b |\psi_2\ket)(a\langle\psi_1|+e^{-i\theta}b\langle\psi_2|)\Bigr]  \nonumber\\  \nonumber  
      & = &    a^2 \rho_{J}  +  
           e^{i \theta} ab \trp_{J ^C} \, | \psi_1 \kb \psi_2 | +  
             e^{- i \theta} ab \trp_{J ^C}  \, | \psi_2 \kb \psi_1 |   + b^2 \rho_{J}   \\
              & = &  \rho_{J}   + ab \Big(  e^{i \theta}  \,  \trp_{J ^C}\, | \psi_1 \kb \psi_2 | + 
               e^{- i \theta}  \,  \trp_{J ^C}\, | \psi_2 \kb \psi_1 |  \Big)
 \ee
 which implies 
 \be
    0 =  e^{i \theta}  \,  \trp_{J ^C} \, | \psi_1 \kb \psi_2 | +  e^{- i \theta}  \,  \trp_{J ^C} \, | \psi_2 \kb \psi_1 |  
 \ee
 which is equivalent to  $e^{i 2 \theta}  A =  A^\dag$ when $A =    \trp_{J ^C} \, | \psi_1 \kb \psi_2 |$. Since this holds for $\theta $ arbitrary, $A =  \trp_{J^C} \, | \psi_1 \kb \psi_2 | = 0 $.

To prove (ii) it suffices to observe that (i) implies
 \be
   \bra \psi_1,  B_{J} \psi_2 \ket  = \trp_{J}  B_{J}  \trp_{J^C}
        | \psi_1 \kb \psi_2 | = \trp_{J} B_{J} 0 = 0.
 \ee
 
 Erdahl applied this Lemma to exposed points with $\cG$ the ground state eigenspace of an $m$-local Hamiltonian.  For arbitrary extreme points  the existence of $\cG$ follows from the fact that the pre-image of a face is always a face.  For completeness, we sketch an elementary proof for extreme points.
 \begin{lemma}   \label{lemm:extalt}
Let   $  |\psi_1 \ket ,  |\psi_2 \ket $ be a pair of orthogonal vectors in $\cH$
 such that  $\trp_{J^C} \proj{\psi_1} = \tr_{J^C} \proj{\psi_2} = \rho_J $ and assume that $\rho_J$ is an extreme point  of
$  \cD_J(\cH_N) $. % \equiv  \hbox{ convex hull } \{ \trp_{J^C} \proj{\psi} : \psi \in \cH_N \} .$$
Then $ \trp_{J^C} \proj{\psi} = \rho_J $ for any unit vector $|\psi\ket \in \text{span} \{|\psi_1\ket, |\psi_2\ket \} $.
\end{lemma} 
\noindent{\bf Proof:}  Let $a, b \in   {\bf C} $, satisfy  $ |a|^2 + |b|^2 = 1 $ and define
     $| \psi_{\pm}  \ket = a | \psi_1\ket    \pm  b | \psi_2\ket  $.   
Then it is easy to verify that 
 \be
       \half \tr_{J^C} \proj{\psi_+} + \half  \tr_{J^C}  \proj{\psi_-} =  \rho_J
 \ee
contradicting the assumption that $\rho_J$ is extreme unless 
      $ \tr_{J^C} \proj{\psi_+}  =  \tr_{J^C}  \proj{\psi_-} = \rho_J $ which implies
 $$          a \ovb{b} \trp_{J^C} | \psi_+ \kb \psi_- | + \ovb{a} b  \trp_{J^C} | \psi_- \kb \psi_+ |  = 0 $$
 which implies that $|\psi_{\pm} \ket  \in \text{span} \{ |\psi_1\ket, |\psi_2\ket \} $.    {\bf QED}

 Note  that since $(a, b)$ is an arbitrary pair of complex numbers with $|a|^2 + |b|^2 = 1$, we can further conclude that  (as above)  $ \trp_{J^C} | \psi_+ \kb \psi_- |  = 0 $.

\begin{thm}   \label{thm:NmON}   {  \rm (Erdahl)}
Let $\{| \psi_k\rangle \}$ be an orthonormal basis for $\cG$ with common $m$-body RDM  $\rho_{J}$ as in  Lemma~\ref{lemma:key}.  Then for any pair with $j \neq k$, 
 $\ran( \trp_{J} \proj{\psi_j })  \subseteq \ker  (\trp_{J} \proj{\psi_k}) $ or, equivalently in the case $J = \{ 1, 2, \ldots ,m \} $,
\be      \label{SchON}
          |\psi_j \ket  & = & \sum_t \mu_t |\chi_t \ot \theta_t^j \ket  
\ee  
with $\mu_t$ and $\{|\chi_t \ket \}$the eigenvalues and eigenvectors of $\rho_m$ and $\bra \theta_s^j, \theta_t^k \ket =  \delta_{jk} $ for all $s, t$.
\end{thm}  
\noindent{\bf Proof:}  The expansion \eqref{SchON} is the standard Schmidt decomposition with  $\{|\theta_t^j \ket\}$  the eigenvectors of $\trp_{J^C} \proj{\psi_j }$. There is no loss of generality in assuming that $\mu_t > 0 $.  By Lemma~\ref{lemma:key}(i)
 \be   \label{iii}
    0  & = &  \trp_{J^C}   | \psi_j \kb \psi_k |  \nonumber \\  \nonumber  
      & = &  \sum_s \sum_t  \mu_s \mu_t 
         \trp_{J^C}   | \chi_s  \ot  \theta_s^j \kb   \chi_t  \ot  \theta_t^k | \\
          & = & \sum_s \sum_t  \mu_s \mu_t   \,   \bra    \theta_s^j, \theta_t^k \ket   \,  | \chi_s \kb  \chi_t| .
 \ee 
Since the set $\{    | \chi_s \kb  \chi_t |    \} $ is  an orthonormal basis for $\cB(\cH^{\ot m}) $ with respect to the Hilbert-Schmidt inner product, the coefficients in \eqref{iii} above must be zero.   By assumption, the singular values $\mu_s \mu_t  > 0$ are all non-zero.   Therefore
\be \label{ortho}
     \bra    \theta_s^j, \theta_t^k \ket=0  \quad \forall ~~s, t  \qquad \hbox{when} ~~ j \neq k.
\ee 
This gives \eqref{SchON} and implies  $\ran (\trp_{J} \proj{\psi_j })  \subseteq  \big(\ran  (\trp_{J} \proj{\psi_k}) \big)^\perp$. The result then follows from $\ker A = (\ran A)^\perp $ for any self-adjoint $A$.     ~~{\bf QED}

To see how remarkable Theorem~\ref{thm:NmON} is,  consider the case $N= 5$, $m = 2$ and suppose that  $| \psi_1 \ket , |\psi_2 \ket $  have the same 2-body RDM.  Then \eqref{SchON} becomes
\bee
     | \psi_1 \ket = \sum_t  \mu_t  \,  | \chi_t(1,2) \ot \theta_t(3,4,5)    \ket \\
       | \psi_2 \ket = \sum_t  \mu_t  \,  | \chi_t(1,2) \ot \phi_t(3,4,5)    \ket 
\eee
with $ \{ | \theta_t  \ket \} $ and  $ \{ | \phi_t  \ket \} $ spanning orthogonal subspaces of $\cH^{\ot 3}$. However, 
\be
      \sum_t |\mu_t|^2  \trp_3 \proj{\theta_t} =   \rho_{45} =   \sum_t |\mu_t|^2  \trp_3 \proj{\phi_t}
\ee
so that the convex hulls of the 2-body RDM associated with these orthogonal subspaces are identical.
% Is it reasonable to conjecture that the common RDM is
%extreme if and only if these sets overlap at a single point??

\subsection{Pre-images of extreme points}\label{sub:ext}

When an extreme point is exposed, it follows that the pre-image is
the convex hull of unit ball of the ground state eigenspace so that the
hypothesis of Theorem~\ref{thm:NmON} holds. Lemma~\ref{lemm:extalt} implies that Theorem~\ref{thm:NmON} also holds for arbitrary extreme points, enabling us to prove  Theorem~\ref{thm:unique}.
 
\noindent{\bf Proof of Theorem~\ref{thm:unique}:}  This is  an easy corollary of Theorem~\ref{thm:NmON}. Let $\cG_J$,   $\cG_{J^c} $ be the subspaces of $ \cH_N$ with common RDM $\rho_J$ and  $\rho_{J^c}$ respectively. $\cG = \cG_J \cap \cG_{J^c}$.  By assumption  $\cG $ is not empty.  If $\dim \cG \geq 2$ then it contains a pair of vectors satisfying Theorem~\ref{thm:NmON}.  But then 
$ \rho_{J^C} = \trp_J \proj{ \psi_1 } =  \trp_J \proj{ \psi_2 } $ which implies that $\theta_s^j = \theta_s^k $ contradicting the strong orthogonality condition \eqref{ortho}.
When $\cH_N$ is the symmetric or anti-symmetric subspace of $\cH^{\ot N} $, and $2m \geq N $, then the same argument can be applied with $J = \{ 1, 2, \ldots, m \}$ and $J^C = \{ m+1, \ldots N \}$.    \qed
 
In the anti-symmetric case, particle-hole duality gives the following
\begin{cor}
For any $N$-fermion system, any extreme point of $\cD^m(\cH^{\wedge N})$ has a unique pre-image if $2m\geq d-N$.
\end{cor} 

In view of Theorem~\ref{thm:unique}, and experience with situations encountered in atomic and molecular problems, Erdahl conjectured that 
the pre-image of every extreme point of $\cD^m(\cH^{\wedge N}) $ is unique. 

In \cite{Ocketal}, it was shown that this conjecture is false, even in the case 
of exposed points, for both the original fermionic situation  and the quantum marginal problem.  As observed in \cite{Ocketal}, Lemma~\ref{lemma:key} implies that a subspace of that form is a quantum error correcting code (QECC) which can correct $\lfloor \frac{m-1}{2}\rfloor$ errors or detect $m - 1$ errors. Although most QECC are associated with interior rather than extreme points of the $m$-body RDM, examples of QECC which arise as the ground state eigenspace of an $m$-local Hamiltonian are now known.

These codes are typically associated with spin systems.   A standard second quantization argument maps the spin eigenstates to fermionic states with only half-filled shells.   To ensure that these are the ground states, the Hamiltonian includes a ``penalty term'' of the form used in the Hubbard model.  The classic examples of extreme points of 2-body RDM exhibit strong pairing, as the case of  ``tight binding'' and unique pre-images.  
%The non-unique pre-images presented in \cite{Ocketal} correspond to the  Mott transition in which the Hamiltonian is decomposable.
 %\bigskip
 
There are presumably large families of extreme points intermediate between this cases, which are yet to be found.

\subsection{Pre-images of non-extreme points}\label{sub:mix}

Since  $\dim  \cH^{\ot N} $ is much larger than $ \dim  \cH^{\ot m} $, the partial trace operation will, in general, take many distinct states to the same output state.
Intuitively, this suggests that non-extreme points of $\cD^m(\cH_N)$ to have multiple pre-images.   Although this is the generic situation, there are some significant
exceptions which we  discuss before stating a result about non-uniqueness. Building on notation introduced previously, (e.g., the pair   $(\rho_J, \rho_{J^\prime}$ )
in Theorem~\ref{thm:diosi}) we let $\vec{Q} = (\rho_{J_1} , \rho_{J_2}  ,   \ldots   ,\rho_{J_\nu} )$ denote a vector of RDM and $\cD_{J_1, J_2, \ldots ,J_\nu}(X) $ the convex set 
of such vectors with pre-images  in $X$.   We are interested in  the conditions under which $\vec{Q}$ does or does not have a unique   pre-image  in $\cD(\cH_N)$, and let 
$\Gamma :     \cD(\cH_N) \mapsto    \cD_{J_1, J_2, \ldots, J_\nu}(\cH_N ) $ for which  $\Gamma(\rho) = \vec{Q}$ (where we have suppressed the dependence on $J_i$ for  simplicity).  

If a  RDM vector $\vec{Q}$ does not lie on a face of  $\cD_{J_1, J_2, \ldots, J_\nu}(\cH_N) $ which is also simplex, it will have multiple decompositions into extreme points which we can write $\vec{Q} = \sum_k a_k \vec{Q_k}  = \sum_k a_k^\prime  \vec{Q_k^\prime} $ and assume  that each  $\vec{Q_k}$ or $\vec{Q_k^\prime} $ has a  unique pre-image  $\proj{\psi_k}$ or $\proj{\psi_k^{\prime}}$, respectively. Then  both  $\rho \equiv \sum_k  a_k \proj{\psi_k} $ and $\rho^\prime \equiv \sum_k  a_k^\prime \proj{\psi_k^\prime} $ are pre-images of  $\vec{Q}$ which suggests non-uniqueness.    However, it is possible that the two states $\rho  = \rho^\prime$ despite the formally distinct decompositions.   An example of such a  situation occurs when an $m$-local Hamiltonian has a doubly degenerate ground state  eigenspace; then  the corresponding set of $N$-body RDM for the ground state is  isomorphic to the Bloch ball.  The image of the Bloch ball under partial trace is also an ellipsoid which corresponds to a face of $\cD^m(\cH_N)$.   Each element of this face, even those which are not extreme, has a unique (possibly mixed) state pre-image. This example also describes a situation in which the map from $\Gamma: \cD^N(\cH_N)  \mapsto \cD^m(\cH_N)$ is one-to-one.   One simply chooses the subspace $\cH_N$ to be the ground state eigenspace of such an $m$-local Hamiltonian with degeneracy.

 A less artificial  example of a one-to-one mapping occurs when $\cH_N = \cH^{\wedge N} $ is  anti-symmetric  and  $\dim \cH = d = N + m$ so that  $\dim \cH^{\wedge N} = \binom{d}{N} =   \binom{d}{m}  = \dim \cH^{\wedge m} $. Then   $\Gamma: \rho^N \mapsto \rho^m$ is given by particle-hole duality as described in \cite[Section 4]{E} and \cite{MBR_PH}.   Moreover, the maps from $\rho^N \mapsto \rho^{\wtd{m}}$ must also be one-to-one for $\wtd{m} > d - N $.

In view of these examples, the following theorem seems to be best possible. For the anti-symmetric case, a stronger result was claimed by Rosina \cite{Rosina}, who apparently did not realize that  the kernel of $\Gamma$ could be trivial.   His argument was similar to that given below.   Since
\be
     \ker (\Gamma) = \{ T  \in \cB(\cH_N)  :    \trp_{J_i^c} T = 0, \quad    i = 1, 2 \ldots \nu \},
\ee
it follows that $\tr T = 0$.   % and $T \in  \ker (\Gamma)  \Leftrightarrow  T^\dag  \in  \ker (\Gamma)$.
Moreover, a non-trivial kernel  contains at least one self-adjoint matrix which we can write as $T = T_+ - T_- $ with both $T_+$ and $T_-$ positive semi-definite.  For any $\rho$ of  full rank and sufficiently small $\mu$,   the matrix $\rho + \mu T \in \cD(\cH_N) $ and $\Gamma(\rho + \mu T ) = \Gamma(\rho) $.   Thus, $\Gamma$ has a non-trivial kernel if and only if $\cD_{J_1, J_2, \ldots, J_\nu}(\cH_N) $ contains two distinct density matrices with the same pre-image.

  \begin{thm}\label{thm:kernel}
If the map $\Gamma:  \rho   \mapsto \vec{Q}$ has a non-trivial kernel, then all interior points of   $\cD_{J_1, J_2, \ldots, J_\nu}(\cH_N) $ have non-unique pre-images in $\cD(\cH_N)$.
Moreover, in this situation, every interior point has at least two pre-images on the boundary of $\cD(\cH_N)$.
\end{thm}
\pf    Let $\vec{Q}$ be an interior point of   $\cD_{J_1, J_2, \ldots, J_\nu}(\cH_N) $,
Since   $\Gamma$ is linear\footnote{Here we use the fact that a linear map is continuous and interpret open as relatively open.}   the inverse image of any ball around $\vec{Q}$ is open.   Thus, there is a point $\vec{Q}^\prime $ in the interior which has a pre-image  $\rho^\prime $ in the interior of $\cD(\cH_N) $. We can find $\eps \in (0,1) $ and $\rho^{\prime \prime} \in  \cD_{J_1, J_2, \ldots, J_\nu}(\cH_N) $ such that $\rho = (1 - \eps )\rho^\prime + \eps \rho^{\prime \prime} $.  Then the pre-images satisfy $\norm{ \rho- \rho^\prime} = \eps \norm{ \rho^\prime - \rho^{\prime \prime}}  \leq 2 \eps$. By choosing a sufficiently small ball around $\rho$ one can make $\eps$ arbitrarily small so that $\rho$, the pre-image of $\rho$, is in the interior of $\cD(\cH_N) $.

To prove the second part, we need the fact that every interior point of $\cD(\cH_N) $ has full rank and consider the line  $\rho + \mu T$ with $T \in \ker(\Gamma)$.   Then $\Gamma (\rho + \mu T ) = \Gamma(\rho) = \vec{Q} $ for all $\mu$.  Since $\rho$ has full rank,  $\rho + \mu T$  is an interior of $\cD(\cH_N) $ for sufficiently small $\mu$.   But since $T$ is not positive semi-definite, the line  $\rho + \mu T$   must eventually cross the boundary  of $\cD(\cH_N) $ for some $\mu > 0$ and some $\mu < 0 $.

In contrast to the anti-symmetric case, the symmetric subspace of  $\cH^{\ot N} $ always has a non-trivial kernel for any choice of  $\vec{Q}$  because all RDM for the pair of GHZ   states  $ | 0\cdots 0 \ket  \pm    | 1\cdots 1 \ket   $  are identical.

Now recall that faces of  $\cD(\cH_N) $ correspond to subspaces of $\cH_N$. One can apply Theorem~\ref{thm:kernel} with $\cH_N$ replaced by a subspace  $\cH^\prime_N \subset \cH_N$  to obtain the following corollary.
\begin{cor} \label{cor:face}
If the map $\Gamma :  \rho   \mapsto \vec{Q}$ restricted to  a face  $\cF$ of $\cD(\cH_N) $ has a non-trivial kernel, then all interior points of   $\cD_{J_1, J_2, \ldots, J_\nu}(\cF) $ have non-unique pre-images in $\cF$. Moreover,  every interior point of  $\cD_{J_1, J_2, \ldots, J_\nu}(\cF)$  has at least two pre-images on the boundary of $\cF$.
\end{cor}

When  $\Gamma$ maps two distinct pure states into the same $\vec{Q}$, the line connecting them is also mapped to $\vec{Q}$.    Thus, if the RDM of a pure state do not determine it uniquely among all pure states, then there will always be mixed states with the same image.

 This leaves the following question.   Suppose that   $\proj{\psi}$  is the only pure state mapped to a RDM vector $\vec{Q}$.   Can there also be a mixed state $\rho$  which is mapped to $\vec{Q}$?   By the discussion above, there must then be a line from  $\proj{\psi}$ to a state  $\rho^\prime$ on the boundary which collapses to the single point $\vec{Q}$ under $\Gamma$.  By assumption, $\rho^\prime$ can not be pure, but lies on a face $\cF$ of    $\cD(\cH_N) $. Only in exceptional circumstances will $\Gamma$ be one-to-one on $\cF$.    Then by applying Corollary~\ref{cor:face}, one can find two additional mixed states $\rho^{\prime \prime}$ and   $\rho^{\prime \prime \prime}$ which lie on the boundary of $\cF$.   Continuing this process inductively, we find that the pre-image of $\vec{Q}$ must be a cone whose apex is $\proj{\psi}$ and for which typical extremal rays go from $\proj{\psi}$ to a mixed state
 on a one-dimensional face (i.e., a line) on the boundary of $\cD(\cH_N) $. Thus, in general, if some mixed state  has the same image as $\proj{\psi}$, there will be a mixed state of rank two with the same image.  An equivalent characterization of rank two mixed states is that they are the reduction of pure state after extension to a qubit environment.\footnote{This fits into our framework using a subspace of $\cH^{ \ot (N+1)} $ for which $\cH_{N+1}  \ot  \subseteq  \cH^{\ot N} \ot {\bf C}_2$.}

In a series of papers, \cite{LPW02,LPW02B,JL05}   Linden, et all showed that the answer to the question above  is negative in certain special situations, including a triplet of qubits.    Their main focus in the later papers is on the number of elements in $\vec{R}_m$ needed to determine the state when the number of parties $N$ is large.    Their work uses parametric analysis so that ``generic''  can be interpreted as   ``almost every'' in a probabilistic setting. Although settings with a trivial kernel seem extremely rare, we do not have a similar measure theoretic interpretation.

\section{Summary and open questions}\label{sect:sum}

Let us summarize the various situations we know
\begin{itemize}

\item[a)]  Almost all pure states are uniquely determined by a pair of  RDMs  $\rho_J, \rho_J^\prime$ which overlap in the sense that  $ J \cap J^\prime \neq \O$ and $J \cup J^\prime = \{ 1, 2, \ldots N \}$.
In the language of \cite{diosi,LPW02,LPW02B} we can say that almost all pure
state are uniquely determined by a pair of RDM $\rho_{JK}, \rho_{KL} $ where $J \cup K \cup L = \{ 1,2, \ldots ,N \}$;
  
\item[b)]  When both $(\rho_J$ and $\rho_{J^C} )$  are extreme points
  of the set of RDM $\cD_J(\cH_N)$ and $\cD_{J^C}(\cH_N) $
  respectively,  there is a unique pure state pre-image satisfying $
  \trp_{J^C} \proj{\psi} = \rho_J $ and $ \trp_J \proj{\psi} = \rho_{J^C} $;
   
\item[c)]  When $2m \geq N$ and $\rho_m \equiv \rho_{1,2, \ldots ,m}$ is an extreme point of the set of symmetric or anti-symmetric $N$-representable $m$-body RDM, it has a unique pre-image;
  
\item[d)]  When $2m < N $,  there are extreme points of the set of $m$-body consistent quantum marginals $\vec{R}_m$  with non-unique pre-image.   These map to extreme points of the set of $m$-body fermionic RDM with non-unique pre-image.  For $m = 2$, the smallest $N$ for which an example is known is $N = 9 $.  

\item[e)]  If the contraction map $\Gamma :     \cD(\cH_N) \mapsto    \cD_{J_1, J_2, \ldots, J_\nu}(\cH_N ) $  has a non-trivial kernel, then all interior points of   $\cD_{J_1, J_2, \ldots, J_\nu}(\cH_N) $ have non-unique pre-images in $\cD(\cH_N)$. Moreover, in this situation, every interior point has at least two pre-images on the boundary of $\cD(\cH_N)$.

\end{itemize}
 
A number of open questions remain.
 
\begin{itemize} 

\item[A)]   Is there an extreme point of the set of $N$-representable $m$-body RDM or consistent $m$-body quantum marginals which is not exposed? 

\item[B)]  What is the smallest $N$ for which an extreme point of the set of  2-body consistent quantum marginals has a non-unique pre-image. Known results on QECC imply $N\geq 5$\cite{Calderbank:1998wy}.

\item[C)]  If  there is only one pure state with RDM vector $\vec{Q}$, can there be a mixed state with the same RDM vector $\vec{Q}$.  We showed in the previous section that for generic situations, any such mixed state has rank two or, equivalently,  it suffices to consider qubit environments.

\end{itemize}

The question of whether or not  every face of the set of $m$-body reduced density matrices is exposed is of some theoretical interest. On the other hand, the relevance for the $N$-representability problem is not so clear.   As pointed out by Kummer \cite{Kum} the set of $m$-body $N$-representable RDM is the closed convex hull of its exposed points  
so that any extreme point which is not exposed is arbitrarily close to an exposed point. Moreover, for the  original motivation of a constrained variational computation for an $N$-body Hamiltonian,  it clearly suffices to consider the exposed points.

Finally we remark that we use $\cH_N\subseteq \cH^{\ot N}$  because that is the situation most commonly considered. However, most of our results apply more generally to subspaces of  $\cH^1 \ot  \cH^2  \ot  \cdots   \ot \cH^N  $  because such situations can be embedded as a subspace of some  $\cH^{\ot N}$ .

\section{Acknowledgement}
JC is supported by NSERC and NSF of China (Grant No. 61179030). ZJ acknowledges support from NSERC and ARO. MBR is supported by US NSF. BZ is supported by NSERC and CIFAR. DZ is supported by NSF of China (Grant Nos. 10975181 and 11175247), and NKBRSF of China (Grant No. 2012CB922104).

\bibliographystyle{plain}
\bibliography{ErdCmmt}

\begin{thebibliography}{10}

\bibitem{KlyCMP}
M.~{Altunbulak} and A.~{Klyachko}.
\newblock {The Pauli principle revisited}.
\newblock {\em Comm. Math. Phys.}, 282:287--322, Sep. 2008.

\bibitem{Calderbank:1998wy}
A.~Robert Calderbank, Eric~M. Rains, P.~W. Shor, and Neil J.~A. Sloane.
\newblock {Quantum error correction via codes over GF(4)}.
\newblock {\em IEEE Trans. Inform. Theory}, 44(4):1369--1387, 1998.

\bibitem{CM}
M.~{Christandl} and G.~{Mitchison}.
\newblock {The spectra of quantum states and the Kronecker coefficients of the
  symmetric group}.
\newblock {\em Comm. Math. Phys.}, 261:789--797, Feb. 2006.

\bibitem{Cole}
A.~J. {Coleman}.
\newblock {Structure of fermion density matrices}.
\newblock {\em Rev. Mod. Phys.}, 35:668--686, Jul. 1963.

\bibitem{diosi}
L.~{Di{\'o}si}.
\newblock {Three-party pure quantum states are determined by two two-party
  reduced states}.
\newblock {\em Phys. Rev. A}, 70(1):010302, Jul. 2004.

\bibitem{E}
R.~M. {Erdahl}.
\newblock {The convex structure of the set of $N$-representable reduced
  $2$-Matrices}.
\newblock {\em J. Math. Phys.}, 13:1608--1621, Oct. 1972.

\bibitem{GP}
C.~{Garrod} and J.~K. {Percus}.
\newblock {Reduction of the $N$-particle variational problem}.
\newblock {\em J. Math. Phys.}, 5:1756--1776, Dec. 1964.

\bibitem{JL05}
N.~S. Jones and N.~Linden.
\newblock Parts of quantum states.
\newblock {\em Phys. Rev. A}, 71:012324, Jan. 2005.

\bibitem{Kly0}
A.~{Klyachko}.
\newblock {Quantum marginal problem and representations of the symmetric
  group}.
\newblock arXiv:quant-ph/0409113, Sep. 2004.

\bibitem{Kly}
A.~{Klyachko}.
\newblock {Quantum marginal problem and $N$-representability}.
\newblock {\em J. Phys.: Conf. Ser.}, 36:72--86, Apr. 2006.

\bibitem{Kuhn}
H.~Kuhn.
\newblock Linear inequalities and the pauli principle.
\newblock {\em Proc. Symp. Appl. Math.}, 10:141--147, 1960.

\bibitem{Kum}
H.~{Kummer}.
\newblock {$N$-representability problem for reduced density matrices}.
\newblock {\em J. Math. Phys.}, 8:2063--2081, Oct. 1967.

\bibitem{LPW02}
N.~Linden, S.~Popescu, and W.~K. Wootters.
\newblock Almost every pure state of three qubits is completely determined by
  its two-particle reduced density matrices.
\newblock {\em Phys. Rev. Lett.}, 89:207901, Oct. 2002.

\bibitem{LPW02B}
N.~Linden and W.~K. Wootters.
\newblock The parts determine the whole in a generic pure quantum state.
\newblock {\em Phys. Rev. Lett.}, 89:277906, Dec. 2002.

\bibitem{QMA}
Y.-K. {Liu}, M.~{Christandl}, and F.~{Verstraete}.
\newblock {Quantum computational complexity of the $N$-representability
  problem: QMA complete}.
\newblock {\em Phys. Rev. Lett.}, 98(11):110503, Mar. 2007.

\bibitem{Ocketal}
S.~A. {Ocko}, X.~{Chen}, B.~{Zeng}, B.~{Yoshida}, Z.~{Ji}, M.~B. {Ruskai}, and
  I.~L. {Chuang}.
\newblock {Quantum codes give counterexamples to the unique preimage conjecture
  of the $N$-representability problem}.
\newblock {\em Phys. Rev. Lett.}, 106(11):110501, Mar. 2011.

\bibitem{Rosina}
M.~Rosina.
\newblock Some theorems on uniqueness and reconstruction of higher-order
  density matrices.
\newblock In J.~Cioslowski, editor, {\em Many-electron densities and reduced
  density matrices}. Springer, 2000.

\bibitem{MBR_PH}
M.~B. Ruskai.
\newblock {$N$-representability problem: particle-hole equivalence}.
\newblock {\em J. Math. Phys.}, 11(11):3218--3224, 1970.

\bibitem{MBR}
M.~B. {Ruskai}.
\newblock {Connecting $N$-representability to Weyl's problem: the one-particle
  density matrix for $N = 3$ and $R= 6$ }.
\newblock {\em J. Phys. A: Math. Gen.}, 40:F961--F967, Nov. 2007.

\bibitem{smith}
D.~W. {Smith}.
\newblock {$N$-representability problem for fermion density matrices. I. The
  second-order density matrix with $N=3$}.
\newblock {\em \jcp}, 43:258, Jul. 1965.

\bibitem{Yang:1962}
C.~N. Yang.
\newblock Concept of off-diagonal long-range order and the quantum phases of
  liquid he and of superconductors.
\newblock {\em Rev. Mod. Phys.}, 34:694--704, Oct. 1962.

\end{thebibliography}

\end{document}